\RequirePackage{amsmath}
\documentclass[runningheads]{llncs}
\usepackage[T1]{fontenc}
\usepackage{amsmath}
\usepackage{hyperref}
\usepackage[capitalize,noabbrev,nameinlink]{cleveref}
\usepackage{microtype}
\usepackage{tcolorbox}
\usepackage{graphicx}
\usepackage{color}

\newcommand{\anon}{{\mathbf a}}
\newcommand{\guess}{{\mathbf g}}
\newcommand{\wei}{{\mathbf w}}

\begin{document}
\title{History Trees and Their Applications}
%
%
\author{Giovanni Viglietta}
\authorrunning{G. Viglietta}
%
\institute{University of Aizu, Aizuwakamatsu, Japan
\email{viglietta@gmail.com}\\
\url{https://giovanniviglietta.com}}
\maketitle              
\begin{abstract}
In the theoretical study of distributed communication networks, \emph{history trees} are a discrete structure that naturally models the concept that anonymous agents become distinguishable upon receiving different sets of messages from neighboring agents. By conveniently organizing temporal information in a systematic manner, history trees have been instrumental in the development of optimal deterministic algorithms for networks that are both anonymous and dynamically evolving.

This note provides an accessible introduction to history trees, drawing comparisons with more traditional structures found in existing literature and reviewing the latest advancements in the applications of history trees, especially within dynamic networks. Furthermore, it expands the theoretical framework of history trees in new directions, also highlighting several open problems for further investigation.

\end{abstract}
\section{Introduction}\label{s:1}

A \emph{distributed communication network} consists of a finite number of independent computational units known as \emph{agents}, which can send each other messages and modify their internal states based on the messages they receive. These networks generally operate in synchronous or asynchronous \emph{communication steps}, where agents send messages through the links of a (multi)graph, which may be static or dynamic, directed or undirected, connected or disconnected, etc.\footnote{Although this note does not focus on \emph{population protocols}, they can also be modeled within this framework as dynamic communication networks where finite-state agents interact in pairs consisting of an initiator and a responder having different roles.}

A network is \emph{anonymous} if its agents are initially indistinguishable, i.e., they lack unique identifiers and can only be told apart by external input or due to the network's layout. In such networks, it is typically assumed that all agents run the same local \emph{deterministic} algorithm. It is important to note that allowing for randomness would enable agents to generate unique identifiers with high probability, thereby compromising the study's focus on anonymity.

A common algorithmic approach in this setting involves assigning a ``weight'' to each agent, which is then distributed among neighbors at each step according to specific rules. This process is analyzed using standard stochastic methods to understand how these weights eventually converge to a common value~\cite{CL22,KM20}.

Although this ``averaging technique'' enabled the development of general algorithms for anonymous networks, one of its major limitations is its disregard for the network's structural and topological characteristics, providing minimal insight into these aspects. Furthermore, analyzing these algorithms tends to be technically cumbersome, yielding only asymptotic estimates for how quickly the network reaches convergence or stability.

Another line of research, initiated by Angluin~\cite{A80}, explores discrete and algebraic structures, such as the \emph{views} of Yamashita--Kameda~\cite{YK96} and the \emph{graph fibrations} and \emph{minimum bases} of Boldi--Vigna~\cite{BV02}. This approach allows algorithms to fully leverage the network's structure and usually permits a more precise analysis of running times. However, the theoretical frameworks of views and graph fibrations were developed for networks with unchanging topologies, and their successful application has been limited to such static networks.

\emph{History trees} are a newer data structure that inherently includes a temporal dimension and was specifically designed for networks with dynamic topologies. The introduction of history trees has recently led to the development of optimal linear-time algorithms for anonymous dynamic networks~\cite{DV22,DVdisc} and state-of-the-art general algorithms for congested anonymous dynamic networks~\cite{DV23}.

In \cref{s:2}, we outline the basic architecture of history trees, drawing comparisons with Yamashita--Kameda's views and Boldi--Vigna's minimum bases. Additionally, we showcase a straightforward linear-time algorithm for dynamic networks that illustrates the practicality of history trees. In \cref{s:3}, we discuss more advanced applications of history trees in challenging network situations, while also pointing out a number of areas that remain open for investigation.

\section{Basic Structure and Algorithms}\label{s:2}
In this section we focus on anonymous networks operating in synchronous steps, modeled as undirected dynamic multigraphs with no port awareness. Other network models will be discussed in \cref{s:3}.

Throughout this section, the reader may find it beneficial to examine the software available at \url{https://github.com/viglietta/Dynamic-Networks}. The repository includes a simulator that supports the creation and visualization of history trees of user-defined dynamic networks and the testing of fundamental algorithms.

\subsection{History Trees and Construction of a View}\label{s:2.construction}
\begin{figure}[t]
\centering
\includegraphics[scale=0.5]{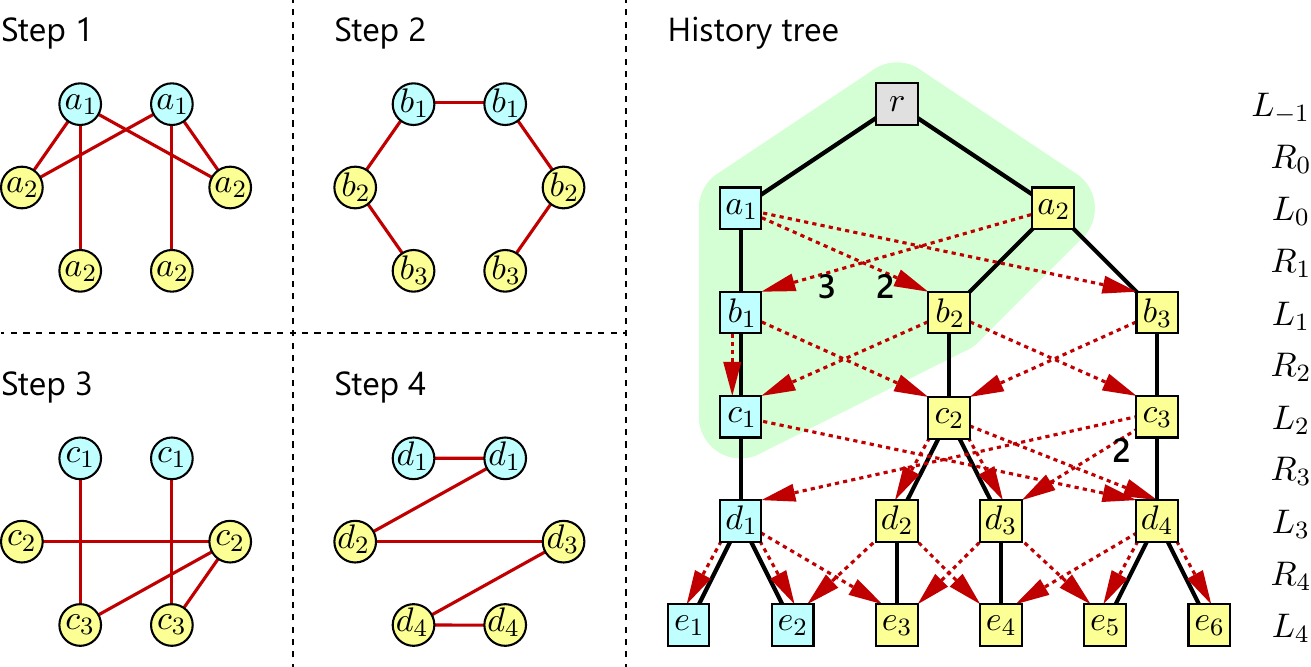}
\caption{The first communication steps of a dynamic network and the corresponding levels of its history tree. Initial inputs are represented by agents' colors (cyan and yellow). Labels on agents and nodes have been added for the reader's convenience and indicate classes of indistinguishable agents. The portion of the history tree with a green background is the view of the two agents labeled $c_1$ after two communication steps.} \label{fig:example}
\end{figure}

Consider a dynamic network of six anonymous agents whose first four communications steps are as in \cref{fig:example}. Before the first step (i.e., at time $t=0$), each agent only knows its own input, which in our example is represented by a color: either cyan or yellow. In the first communication step, both cyan agents (which are indistinguishable) receive three messages from yellow agents (also indistinguishable). Thus, the cyan agents remain indistinguishable at time $t=1$.

However, two yellow agents receive two messages from cyan agents, but the other two receive only one message. Thus, at time $t=1$, the two yellow agents labeled $b_2$ are still indistinguishable from each other, but are distinguishable from the ones labeled $b_3$, because they have a different ``history''.

In general, since the network is synchronous, at step~$t>0$ (i.e., between time $t-1$ and time~$t$) all agents simultaneously send messages and simultaneously receive them (as unordered multisets). Two messages are assumed to be identical if and only if they are sent by indistinguishable agents (possibly by the same agent). By definition, at time $t=-1$ no two agents are distinguishable; at time $t=0$ they are distinguishable if and only if they have different inputs. At time $t>0$, two agents are distinguishable if and only if they are distinguishable at time $t-1$ or receive different multisets of messages at step~$t$. 

\emph{History trees} were introduced in~\cite{DV22} to study how and when anonymous agents in a (dynamic) network become distinguishable. The history tree $\mathcal H_G$ of a network $G$ is an infinite tree subdivided into \emph{levels}, as in \cref{fig:example}, where the nodes in level $L_t$ represent the equivalence classes of agents that are indistinguishable at time~$t$. The number of agents in the class represented by a node $v$ is called the \emph{anonymity} of $v$ and is denoted as $\anon(v)$. In our example, $\anon(a_1)=2$ and $\anon(a_2)=4$.

The root $r$ represents all agents in the network. In general, the \emph{children} of a node $v\in L_{t-1}$, i.e., the nodes $v_1, v_2, \dots, v_k \in L_t$ that are connected to $v$ by a \emph{black edge}, represent a partition of the set of agents represented by $v$. Thus,
\begin{equation}\label{eq:1}
\anon(v) = \sum_{i=1}^k \anon(v_i).
\end{equation}

On the other hand, the \emph{red edges} in $R_t$ represent messages sent and received at step~$t$. That is, a directed red edge $(v,u)$ with \emph{multiplicity} $m$, where $v\in L_{t-1}$ and $u\in L_t$, indicates that, at step~$t$, each agent represented by node $u$ receives exactly $m$ (identical) messages from agents represented by $v$.

Assuming that their internal memory and message sizes are unbounded, the agents in a network $G$ can locally construct portions of the history tree called \emph{views}. More precisely, if an agent $p$ at time $t$ is represented by a node $v\in L_t$, then the view $\mathcal V_G^t(p)$ is defined as the portion of $\mathcal H_G$ spanned by all directed paths from the root $r$ to $v$, using black and red edges indifferently (black edges are assumed to be directed away from the root $r$). The node $v$ is called the \emph{bottom} of the view. \cref{fig:example} shows an example of a view within a history tree.

\begin{figure}[t]
\centering
\includegraphics[scale=0.5]{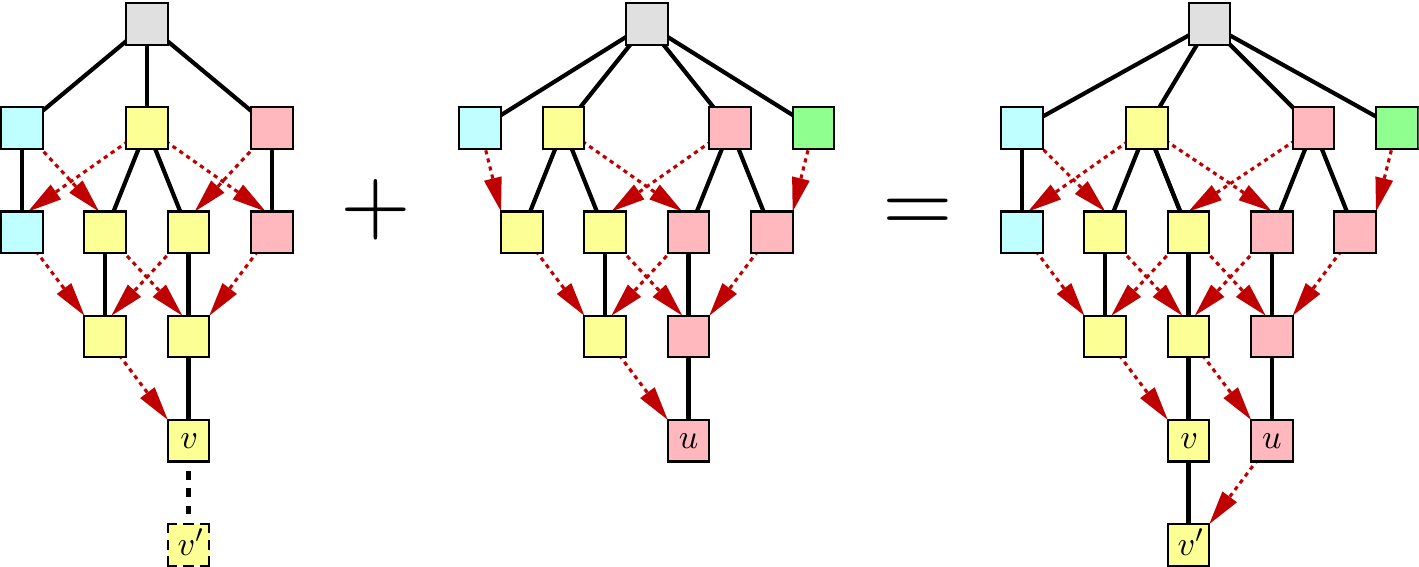}
\caption{Updating the view of an agent represented by node $v$ after it receives the view of an agent represented by $u$ as a message. The two views are matched and merged starting from the roots, $v$ gets a child $v'$, and a new red edge from $u$ to $v'$ is added.} \label{fig:merge}
\end{figure}

The distributed construction of views can be achieved via an iterative process, assuming that all agents send their current view to all their neighbors at every step. Upon receiving a multiset $M$ of messages, an agent $p$ integrates its current view $\mathcal V_G^t(p)$ with all views in $M$. This is done by a straightforward match-and-merge algorithm, starting from the root and proceeding level by level, as exemplified in \cref{fig:merge}. The result is the smallest tree containing $\mathcal V_G^t(p)$ and all views in $M$ as induced subtrees. To obtain $\mathcal V_G^{t+1}(p)$, the agent $p$ creates a child $v'$ for the bottom node $v$ of $\mathcal V_G^t(p)$ and connects the bottom nodes of all views in $M$ to $v'$ by red edges with the appropriate multiplicities. The node $v'$ is now the bottom of $\mathcal V_G^{t+1}(p)$ and represents $p$ (and possibly other agents) at time $t+1$.

It is a simple observation that the view $\mathcal V_G^t(p)$ contains all the information that $p$ can possibly extract from the network $G$ after $t$ communication steps~\cite{DV22}.

\subsection{Related Structures for Static Networks}\label{s:2.related}
We will now focus on \emph{static} anonymous networks, which are well understood thanks to the works of Yamashita--Kameda, Boldi--Vigna, and other authors.

If $G$ is a static network of $n$ agents, once the classes of indistinguishable agents remain unchanged for a single communication step, they must remain unchanged forever. Thus, the number of nodes in the levels of $\mathcal H_G$ strictly increases at every level until it stabilizes, say, at level $L_s$. Therefore, if $|L_s|=\widehat n$, we have $0\leq s< \widehat n\leq n$, because $|L_{0}|\geq 1$ and each node represents at least one agent. 

We come to the conclusion that in any network of size $n$, if two agents are indistinguishable at time $t=n-1$, they will remain indistinguishable thereafter. This observation was first made by Norris (though formulated differently) in~\cite{norris}, where she also raised the question of whether the same applies to agents in two distinct networks of the same size $n$.\footnote{By treating the disjoint union of the two networks as a single network of size $2n$, it becomes evident that agents that are indistinguishable at time $t=2n-1$ are indistinguishable forever. This point was already noted by Conway in~\cite[Chapter~1, Theorem~7]{conway}, albeit phrased in terms of Moore machines. Although $t=2n-1$ might not be an optimal bound if both networks have size $n$ (hence Norris' question), \cref{fig:lower} implies that $t=2n$ is optimal if they have sizes $n$ and $n+1$, respectively. Indeed, if the two cyan agents are colored yellow, they become distinguishable after $2n$ steps, matching the upper bound given by the combined number of agents, $2n+1$.} We restate Norris' question below.

\begin{tcolorbox}
{\bf Open Problem 1.} Let $G$ and $G'$ be two disjoint static networks of $n$ agents each. If an agent in $G$ and an agent in $G'$ have isomorphic views at time~$t=n-1$, do they have isomorphic views at all times?
\end{tcolorbox}

\begin{figure}
\centering
\includegraphics[scale=0.5]{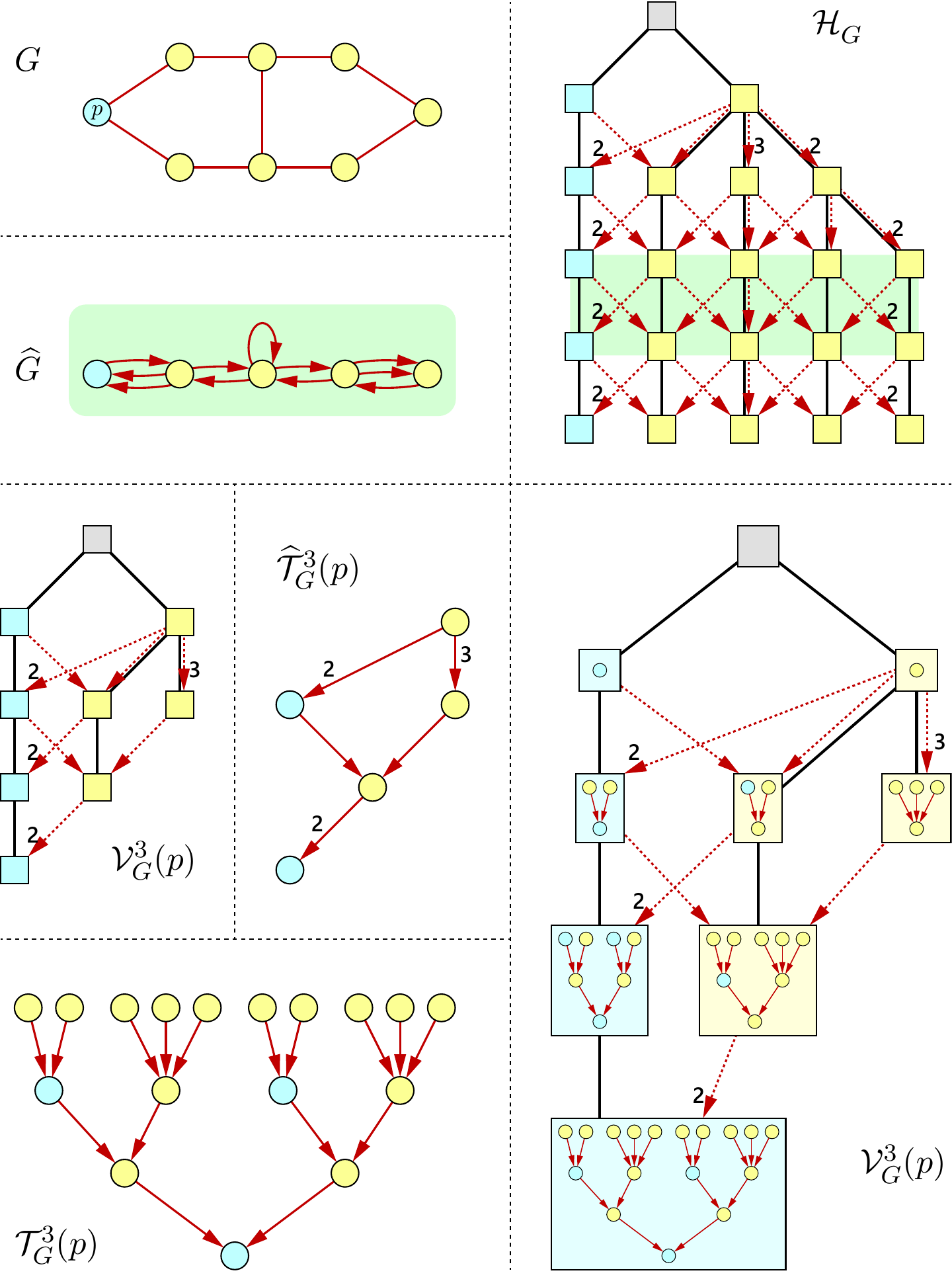}
\caption{A static network $G$ with a distinguished agent $p$ and its history tree $\mathcal H_G$ (truncated at level $L_4$). At stabilization, the directed graph of the red edges between non-branching levels (highlighted in green in $\mathcal H_G$) is isomorphic to the Boldi--Vigna minimum base $\widehat G$. The view $\mathcal V_G^3(p)$ after step~$3$ is related to the Yamashita--Kameda view $\mathcal T_G^3(p)$ truncated at depth~$3$, as well as to its Tani folded view $\widehat{\mathcal T}_G^3(p)$, as illustrated. The precise correspondence between these structures is described in \cref{s:2.related}.} \label{fig:compare}
\end{figure}

Let us consider the subgraph of $\mathcal H_G$ induced by the nodes in levels $L_s$ and $L_{s+1}$, as highlighted in \cref{fig:compare} with a green background. Contracting the black edges in this subgraph, we obtain a directed multigraph named $\widehat G$.

This is an important structure that has been given many names and equivalent definitions in the literature. Essentially, $\widehat G$ could be defined as the smallest directed multigraph that gives rise to a history tree isomorphic to $\mathcal H_G$. Yamashita and Kameda call $\widehat G$ the \emph{quotient graph} of $G$ and define it in graph-theoretical terms using their own notion of \emph{view} of a static network~\cite{YK88,YK96}, whereas Boldi and Vigna call $\widehat G$ the \emph{minimum base} of $G$ and give it a topological definition in terms of \emph{graph fibrations}~\cite{BV96,BV99,BV02,BV02b}.\footnote{More specifically, $G$ is fibered onto the fibration-prime graph $\widehat G$, and the fibers are precisely the classes of agents represented by the nodes of $L_s$ in the history tree $\mathcal H_G$.}

Notably, since the results concerning static networks from these authors are presented with reference to $\widehat G$, they can be directly rephrased in the language of history trees. The advantage of using history trees lies in their ability to readily offer timing information on when agents become distinguishable. This feature makes history trees particularly suitable for dynamic networks, where $\widehat G$ is not well defined and the lack of topological regularity prevents the straightforward deduction of temporal data.

Before delving into applications of history trees to dynamic networks in \cref{s:2.appl}, let us draw a closer comparison between views of history trees and views in the sense of Yamashita--Kameda~\cite{YK96}. While it is important not to confuse these two concepts, they bear a deep relationship illustrated in \cref{fig:compare}.

For Yamashita and Kameda, the view $\mathcal T_G(p)$ of an agent $p$ in a static network $G$ is an infinite tree rooted at $p$, where each node of depth $k$ represents a distinct walk of length $k$ in $G$ terminating in $p$. As it turns out, $\mathcal T_G(p)$ truncated at depth $k$, denoted as $\mathcal T^k_G(p)$, contains precisely the information encoded in $\mathcal V_G^k(p)$.

Indeed, starting from $\mathcal V_G^k(p)$, we can construct $\mathcal T^k_G(p)$ as follows. Let us consider the subgraph spanned by the directed paths in $\mathcal V_G^k(p)$ terminating in the bottom node using only red edges. Such a subgraph is a directed acyclic graph isomorphic to the so-called \emph{folded view} $\widehat{\mathcal T}_G^k(p)$. This is a structure devised by Tani in~\cite{tani} to compress $\mathcal T^k_G(p)$ from exponential size to polynomial size by conflating equivalent nodes. Finally, $\widehat{\mathcal T}_G^k(p)$ can easily be unraveled to reconstruct $\mathcal T^k_G(p)$.

Conversely, starting from $\mathcal T^k_G(p)$, one can construct an inventory of ``fragments'' by listing all subtrees hanging from different nodes and truncating them at all possible depths. The isomorphism classes of these fragments, when ordered by their height, correspond to the levels of $\mathcal V_G^k(p)$, with the ``empty fragment'' being the root. Then, the parent of a fragment $f$ in $\mathcal V_G^k(p)$ is determined by deleting all the leaves of $f$. Similarly, the fragments of $\mathcal V_G^k(p)$ that are connected to $f$ via red edges (and the multiplicities of such red edges) match the subtrees that dangle from the root's children within $f$. Thus, we can construct $\mathcal V_G^k(p)$ from $\mathcal T_G^k(p)$.

\subsection{Basic Applications to Dynamic Networks}\label{s:2.appl}
We will now describe a general algorithmic technique that can be used to solve a wide range of fundamental problems in networks operating in synchronous steps, modeled as \emph{dynamic} undirected multigraphs that are connected at every step. Albeit being extremely straightforward, this technique achieves optimal running times and matches in efficiency the best algorithms for static networks.

As argued in \cref{s:2.construction}, the agents in a network have a distributed algorithm for constructing their view of the history tree and update it at every communication step. Since an agent's view encodes all the information that the agent can infer from the network, in principle we can reduce any problem about anonymous networks to a problem about views of history trees.

Of course, this is true assuming that all agents have enough internal memory and can send large-enough messages to store their views. After $t$ steps, the size of a view is $O(tn^2\log M)$ bits, where $n$ is the total number of agents and $M$ is the maximum number of messages sent by any agent in a single step. Indeed, such a view has $t$ levels, each of which contains at most $n$ nodes and $n^2$ incoming red edges of multiplicity at most $M$. In particular, if the view construction algorithm runs for a polynomial number of steps, it requires internal memory and messages of polynomial size. For the time being, we will assume that this is not an issue.

The basic technique relies on the following observation. If two nodes $u$ and $v$ in the same level $L_i$ of the history tree are \emph{non-branching}, i.e., they have a unique child $u'$ and $v'$ respectively, and there are red edges $(u,v')$ with multiplicity $m_{u,v'}>0$ and $(v,u')$ with multiplicity $m_{v,u'}>0$, then we can count the number of messages exchanged by the agents represented by $u$ and by $v$ as
\begin{equation}\label{eq:2}
m_{v,u'}\, \anon(u) = m_{u,v'}\, \anon(v)
\end{equation}
because communication links are bidirectional. Thus, since we know the multiplicities of the red edges, we can infer the ratio of the anonymities involved.

Now, if $L_i$ is a level where all nodes are non-branching, and since the network is connected at every step by assumption, we can repeatedly apply \cref{eq:2} to compute the ratios of the anonymities of all nodes in $L_i$. For example, since level $L_1$ in \cref{fig:example} is non-branching, we can easily deduce that $\anon(b_1) = \anon(b_2) = \anon(b_3)$. Then, using \cref{eq:1}, we can extend the computation to all previous levels, and obtain for instance that $2\anon(a_1) = \anon(a_2)$. The reader may apply the same method to the history tree in \cref{fig:compare} to deduce that the ratio between the number of yellow agents and the number of cyan agents is $7$.

Once we have this information, we can solve problems such as \emph{Average Consensus}: if the agents are given input numbers, we can compute the fraction of agents that have each input and use it as a weight to compute the mean input.

If we are given additional information, such as the total number of agents or the number of agents that have a certain input, we can also compute how many agents have each input. For instance, in \cref{fig:example}, if we know that there are two cyan agents, we deduce that there are four yellow ones, because $2\anon(a_1) = \anon(a_2)$.

In particular, if we know that the network contains a unique distinguished agent, typically referred to as a \emph{leader}, then we can solve the \emph{Counting problem}, which asks to compute the total number of agents, $n$. Thus, in \cref{fig:compare}, if we know that $p$ is the unique leader, we can deduce from the history tree that $n=8$.

Let us discuss the efficiency of this method. Since $|L_{-1}|=1$ and the number of nodes per level is at most $n$, it follows that it is sufficient to inspect the history tree up to level $L_{n-1}$ to find a non-branching level and the relative red edges.

A subtler issue is how long it takes before the views of all agents acquire all nodes (and hence all edges) up to $L_{n-1}$, so that all agents can carry out correct computations locally. In \cref{fig:example}, for instance, since level $L_0$ is non-branching in the highlighted view (because node $b_3$ is not in the view), at time~$t=2$ the cyan agents may be deceived and incorrectly deduce that $3\anon(a_1) = 2\anon(a_2)$.

Recall that the network is assumed to be connected at all steps. So, it takes fewer than $n$ steps for information to travel from any agent to any other agent; in other words, the \emph{dynamic diameter} $d$ is at most $n-1$. Indeed, if $k<n$ agents have some information at time $t$, then at least $k+1$ agents have it at time $t+1$.

Hence, it takes at most $d$ steps for a node in the history tree to appear in the views of all agents. In particular, at time~$n+d-1\leq 2n-2$, all agents have the entirety of $L_{n-1}$ in their views, and thus have enough information to do correct computations. For example, in \cref{fig:compare}, knowing all levels of $\mathcal H_G$ up to $L_3$ is enough, but these levels entirely appear in the view of $p$ only at step~$7$. In fact, the diameter of the network $G$ is $d=4$.

This technique yields an algorithm for Average Consensus that \emph{stabilizes} in $2n-2$ steps. That is, if all agents attempt to compute the average of the input values after every step, they may guess it incorrectly for some time, but will be all correct after $2n-2$ steps at the latest. The same upper bound holds for the Counting problem, assuming the presence of a unique leader in the network.

A downside of this method is that it provides no certificate of correctness, and therefore the algorithm is supposed to run indefinitely. However, if $n$ is known to the agents, they can simply count $2n-2$ steps, do all computations on the first $n-1$ levels, and then terminate returning the correct output. Similarly, if an agent knows the dynamic diameter $d$, it can terminate at time~$t$ if its view contains a non-branching level $L_i$ with $i< t-d$, because all levels up to $L_{i+1}$ are guaranteed to be entirely in the view. This occurs by the time $t=n+d-1$.

\begin{figure}[t]
\centering
\includegraphics[scale=0.5]{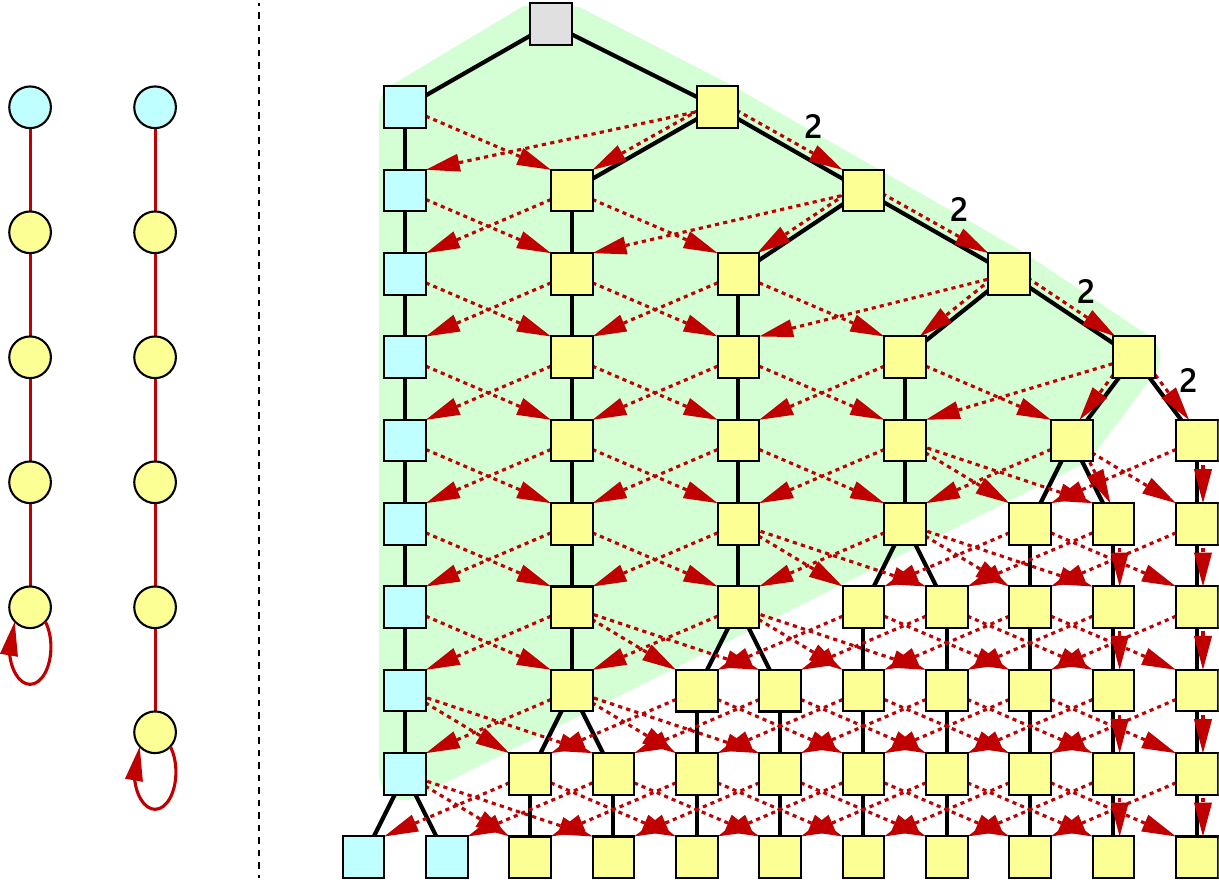}
\caption{A static network with two connected components of sizes $n$ and $n+1$. The history tree shows that the two agents in cyan become distinguishable after $2n-1$ communication steps. Therefore, in static networks of $n$ agents with a unique leader, no Counting algorithm can terminate in $2n-2$ steps or stabilize in $2n-4$ steps.} \label{fig:lower}
\end{figure}

Finally, let us discuss lower bounds. \cref{fig:lower} shows two networks of sizes $n$ and $n+1$ whose leaders have the same view up to step~$2n-2$. Since agents with equal views must return equal outputs (assuming they execute the same algorithm), it easily follows that there is no Counting algorithm that terminates in $2n-2$ steps or stabilizes in $2n-4$ steps in both networks. For if there were such an algorithm, then both leaders would return the same output, and at least one of them would be incorrect. Note that this almost matches the running time of the stabilizing Counting algorithm given above, which is $2n-2$ steps. This leaves only a small gap: namely, whether $2n-3$ steps are actually sufficient.

\begin{tcolorbox}
{\bf Open Problem 2.} Can a Counting algorithm stabilize in $2n-3$ steps in all connected undirected dynamic networks with a unique leader?
\end{tcolorbox}

As for Average Consensus, if the two cyan agents in \cref{fig:lower} are colored yellow, then they have the same view up to step~$2n-1$. This implies that no Average Consensus algorithm can stabilize in $2n-3$ steps in all networks, and therefore the one described above, which stabilizes in $2n-2$ steps, is optimal.

\section{Variations and Extensions}\label{s:3}

\subsection{Leader Election}\label{s:3.leader}
Another application of the technique in \cref{s:2.appl} is \emph{Leader Election}, where all agents have to agree on a unique representative to be identified as the ``leader''. Of course, this problem has a solution only if the history tree contains a node of anonymity~$1$; as it turns out, this condition is also sufficient.

A simple Leader Election algorithm, at step~$t$, computes the ratios of all anonymities in the first non-branching level occurring after $L_{\lfloor t/2\rfloor}$ and deterministically picks a node of smallest anonymity as representing the leader. This strategy eventually elects a unique leader if and only if it is possible to do so, but no certificate of correctness is provided, and the algorithm has to run indefinitely.

However, if $n$ is known, then anonymities in non-branching levels can be computed exactly with a known delay of $n-1$ steps. In this case, any agent that becomes distinguishable from all others at time~$t$ can be identified with certainty by time $t+2n-2$, allowing all agents to terminate.

\subsection{Terminating Counting}\label{s:3.term}
In \cref{s:2.appl} we gave a Counting algorithm for connected undirected dynamic networks with a unique leader that stabilizes in $2n-2$ steps. At the cost of $n$ additional communication steps, we can implement a correctness certificate and make all agents return $n$ and terminate. The algorithm's details are somewhat intricate and can be found in~\cite{DV22}; here we will merely sketch the main ideas.

If we know the anonymities of a node $u$ and of all its children, then $u$ is called a \emph{guesser}. If there is a red edge directed from $u$ to a node $v$ of unknown anonymity, we can write an equation similar to \cref{eq:2} to count messages exchanged by the corresponding agents. Solving this equation yields a \emph{guess} $\guess(v)$ on $\anon(v)$ in terms of known anonymities and multiplicities of red edges. As it turns out, $\guess(v)\geq \anon(v)$ always holds; moreover, if $v$ has no siblings, $\guess(v)= \anon(v)$.

Clearly, the nodes representing the leader are always guessers, because their anonymity is~$1$. Using these nodes, we can make initial guesses on at least one node per level. The question is how we can determine which guesses are correct without knowing the whole history tree, but only a view of it.

Let us define the \emph{weight} $\wei(v)$ of a node $v$ as the number of guesses that have been made on nodes in the subtree hanging from $v$ (including on $v$ itself). Then, $v$ is said to be \emph{heavy} if $\wei(v)\geq \guess(v)$. The key observation is that, if guesses are \emph{well spread}, i.e., no two sibling nodes are assigned a guess, then the deepest heavy node necessarily has a correct guess.

\begin{figure}[t]
\centering
\includegraphics[scale=0.5]{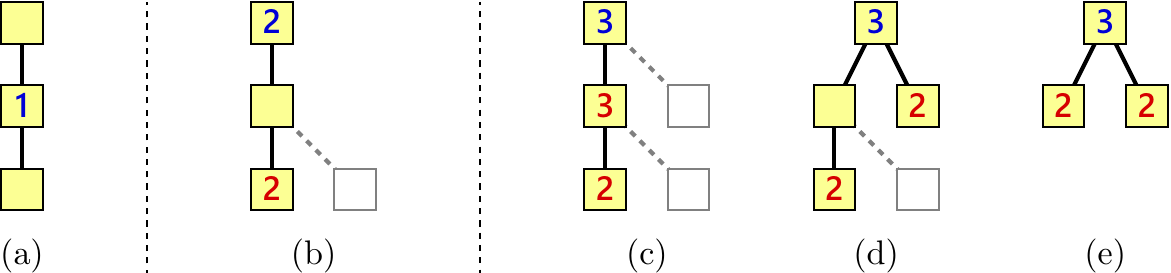}
\caption{The numbers in the nodes are guesses made according to the algorithm in \cref{s:3.term}. Blue numbers indicate heavy nodes; their guesses are necessarily correct except for the one in (e), where guesses are not well spread.} \label{fig:ht2}
\end{figure}

Hence, in \cref{fig:ht2}, the nodes in (a)--(d) with a blue number have a correct guess, but the one in (e) may not. For (a), this is obvious: since $\guess(v)=1$ and $\guess(v)\geq \anon(v)$, we must conclude that $\anon(v)=1$ and the guess is correct. As for (b), if the lower node $u$ has a correct guess, then the upper node $v$ has a correct guess too, because $2=\guess(v)\geq \anon(v)\geq \anon(u) = \guess(u)=2$, hence $\guess(v)= \anon(v)$. Otherwise, the guess on $u$ is incorrect, and so $u$ must have a sibling (perhaps not in the view). Hence, $2=\guess(v)\geq \anon(v)\geq 2$, and the guess on $v$ is correct. The reader may verify that a similar line of reasoning applies to (c) and (d) but not to (e).

It is not difficult to see that, if there are at least $n$ well-spread nodes with guesses, then some of them must be heavy, and thus the deepest of them is necessarily correct. In this way, we can steadily determine the anonymities of new nodes. Eventually, some of these nodes become guessers themselves and can be used to make new guesses, and so on. Since the network is connected at all steps, it can be shown that only $2n-2$ levels of the history tree are required for this algorithm to eventually identify nodes with correct guesses on all branches.

Adding up the anonymities of nodes on all branches produces an estimate $n'$ on the actual size of the network, $n$. In general we have $n'\leq n$, because the current view may not include all branches of the history tree. To confirm this estimate, it is sufficient to wait an additional $n'$ steps, which is the longest time it may take for at least one missing branch to appear in the view, if one exists.

Overall, this Counting algorithm terminates in $3n-2$ steps in the worst case, leaving a small gap with the lower bound of $2n-2$ steps given in \cref{s:2.appl}.
\begin{tcolorbox}
{\bf Open Problem 3.} Can a Counting algorithm terminate in $2n +O(1)$ steps in all connected undirected dynamic networks with a unique leader?
\end{tcolorbox}

The previous Counting algorithm was generalized in~\cite{DVdisc} to networks where the leader may not be unique, but there is a known number $\ell\geq 1$ of indistinguishable leaders, or \emph{supervisors}. Note that the history tree may contain several branches corresponding to leaders; the challenge posed by this scenario is that the individual anonymities of such branches are unknown (although their sum is $\ell$).

The state-of-the-art solution given in~\cite{DVdisc} involves subdividing the history tree into $\ell$ intervals, each of at most $(\ell+1)n$ levels. The total running time is roughly $(\ell^2+\ell+1)n$ steps, making this algorithm impractical unless $\ell$ is small.

In fact, it is a major open question whether a larger $\ell$ may actually be helpful.

\begin{tcolorbox}
{\bf Open Problem 4.} In connected undirected dynamic networks with a known number $\ell$ of leaders, how does the optimal running time of terminating Counting algorithms scale with $\ell$?
\end{tcolorbox}

\subsection{Directed Networks}\label{s:3.directed}
\begin{figure}[t]
\centering
\includegraphics[scale=0.5]{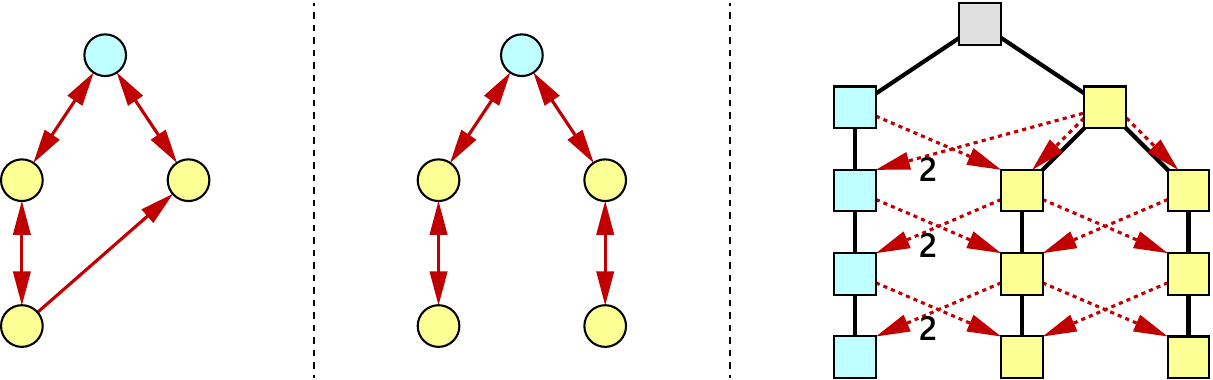}
\caption{Two directed static networks with a unique leader and no outdegree awareness that have a different number of agents but the same history tree.} \label{fig:aware}
\end{figure}

Generalizing the scenario of \cref{s:2}, we may consider \emph{directed networks}, i.e., networks with unidirectional links. It is important to recognize that, in these networks, even solving basic problems such as Average Consensus or Counting with a unique leader requires some additional assumptions, as \cref{fig:aware} demonstrates.

A reasonable approach is to assume that, in each step, agents are aware of their outdegree either before (\emph{early outdegree awareness}) or after (\emph{late outdegree awareness}) they transmit their messages. The key distinction is that in the early awareness model, an agent can incorporate its current outdegree into the messages it sends. Here we will focus on the late outdegree awareness model, which is less advantageous especially in dynamic networks.

\begin{figure}[t]
\centering
\includegraphics[scale=0.5]{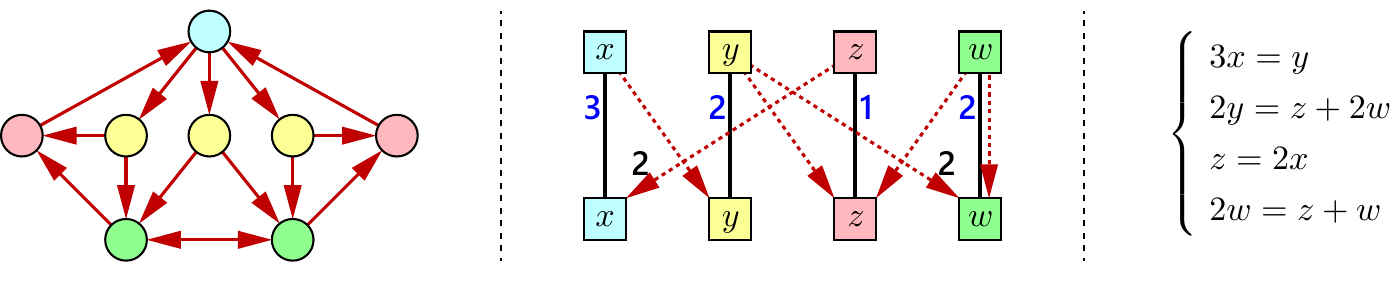}
\caption{A communication step of a directed network with late outdegree awareness and a non-branching level of its history tree. Blue numbers indicate outdegrees.} \label{fig:dir}
\end{figure}

The basic history tree architecture can be extended by attaching outdegrees to black edges, as in \cref{fig:dir}. We can then identify a non-branching level $L_i$ in the history tree, as in \cref{s:2.appl}, and use outdegrees and multiplicities of red edges to write linear equations generalizing \cref{eq:2}, as shown in \cref{fig:dir}.

The resulting homogeneous system is represented by an \emph{irreducible} matrix $A$, assuming the network is strongly connected at every step. Note that $A=\lambda I-P$, where $\lambda >0$ and $P\geq 0$ is also an irreducible matrix. In our example,
$$A=\begin{bmatrix}
3 & -1 & 0 & 0 \\
0 & 2 & -1 & -2 \\
-2 & 0 & 1 & 0 \\
0 & 0 & -1 & 1
\end{bmatrix}
=\begin{bmatrix}
3 & 0 & 0 & 0 \\
0 & 3 & 0 & 0 \\
0 & 0 & 3 & 0 \\
0 & 0 & 0 & 3
\end{bmatrix}
- \begin{bmatrix}
0 & 1 & 0 & 0 \\
0 & 1 & 1 & 2 \\
2 & 0 & 2 & 0 \\
0 & 0 & 1 & 2
\end{bmatrix}
= 3I - P.$$
We have $Ax=0$, where $x>0$ is the vector of anonymities of the nodes in $L_i$. Hence, $Px=\lambda x$. By the Perron-Frobenius theorem, $\lambda$ is a \emph{simple} eigenvalue of $P$, and thus $0$ is a simple eigenvalue of $A$. In other words, the \emph{nullity} of $A$ is~$1$, which means we can solve the linear system in terms of a single free variable.

Thus, we can find the ratios between the anonymities of all nodes in $L_i$, or compute them exactly if there is a known number $\ell\geq 1$ of leaders in the network.

This technique yields stabilizing Average Consensus and Counting algorithms for strongly connected directed dynamic networks with late outdegree awareness, generalizing the ones in \cref{s:2.appl}. The stabilization time of these algorithms is again $2n-2$ steps, which is optimal.

We now present a \emph{terminating} Counting algorithm for directed networks with late outdegree awareness and a known number $\ell\geq 1$ of leaders. Every agent waits until its view contains a long-enough interval $\mathcal I$ of non-branching levels. The goal is to find an upper bound $U_i$ on the anonymity of a node $v_i$ in each branch $B_i$ within $\mathcal I$. To start, we take $U_1=\ell$, where $B_1$ is any leader branch.

The algorithm repeatedly uses branches with a known upper bound to determine new upper bounds. Namely, let $B_i$ have a known upper bound $U_i\geq \anon(v_i)$ and consider all red edges $(v,v')$, where $v$ is a descendant of $v_i$ within $\mathcal I$. Each of these yields an \emph{estimate} $\delta U_i$ on $\anon(v')$, where $\delta$ is the outdegree corresponding to the child of $v$ in $B_i$. Note that, if $v$ has a unique child in the history tree, at most $\delta U_i$ messages are sent from agents represented by $v$, and so $\delta U_i\geq \anon(v')$. Nonetheless, even if $v$ is not branching in a view, it may still branch in the history tree, and in this case $\delta U_i$ may not be a correct upper bound. However, this undesirable event may occur at most $U_i-1$ times. Thus, as soon as a branch $B_j$ receives estimates from $B_i$ on $U_i$ nodes, we take the maximum estimate as a correct upper bound $U_j$ on the anonymity of the \emph{deepest} such node $v_j$ in $B_j$.

When we finally have an upper bound for all branches in $\mathcal I$, we wait $\sum_i U_i$ additional steps to confirm that there are no branches missing from the view. Then we run the previous stabilizing Counting algorithm and output the result.

Assuming that the network is simple and strongly connected at all steps, this algorithm terminates in $2^{O(n \log n)}$ steps, which is likely far from optimal.

\begin{tcolorbox}
{\bf Open Problem 5.} Can a Counting algorithm terminate in a \emph{polynomial} number of steps in all strongly connected directed dynamic simple networks with (early or late) outdegree awareness and a unique leader?
\end{tcolorbox}

\subsection{Disconnected Networks}\label{s:3.disc}
For networks that are not necessarily connected at all steps, we define a \emph{communication round} as a minimal sequence of consecutive steps whose communication multigraphs have a (strongly) connected \emph{sum} (constructed by adding together their adjacency matrices). A network is \emph{$\tau$-union-connected} if every block of $\tau$ consecutive steps contains a communication round. It was observed in~\cite{DVdisc} that $\tau$ and the dynamic diameter $d$ are related by the tight inequalities $\tau\leq d\leq \tau(n-1)$.

Clearly, any non-trivial \emph{terminating} computation is impossible if the agents know nothing about $\tau$. On the other hand, with knowledge of $\tau$, all of the previous algorithms can be straightforwardly adapted to $\tau$-union-connected networks. It is sufficient for each agent to accumulate incoming messages at every step, updating its view only once every $\tau$ steps. This adaptation slows down all running times by a factor of $\tau$. However, this is worst-case optimal, since a $\tau$-union-connected network may be devoid of links except at steps that are multiples of $\tau$.

As for \emph{stabilizing} computations, they can be done even with no knowledge of $\tau$, again with a worst-case optimal slowdown by a factor of $\tau$, as detailed in~\cite{DVdisc}.

\subsection{Semi-Synchronous Networks}\label{s:3.ssynch}
In a network operating \emph{semi-synchronously}, any agent may unpredictably be \emph{inactive} at any step. An inactive agent does not communicate with other agents and does not even update its state. This model is called ``asynchronous'' by Boldi--Vigna~\cite{BV99,BV02b} and generalizes the \emph{asynchronous starts} of Charron-Bost--Moran~\cite{CB19}, but we will give the term ``asynchronous'' a more extensive meaning in \cref{s:3.asynch}.

\begin{figure}[t]
\centering
\includegraphics[scale=0.5]{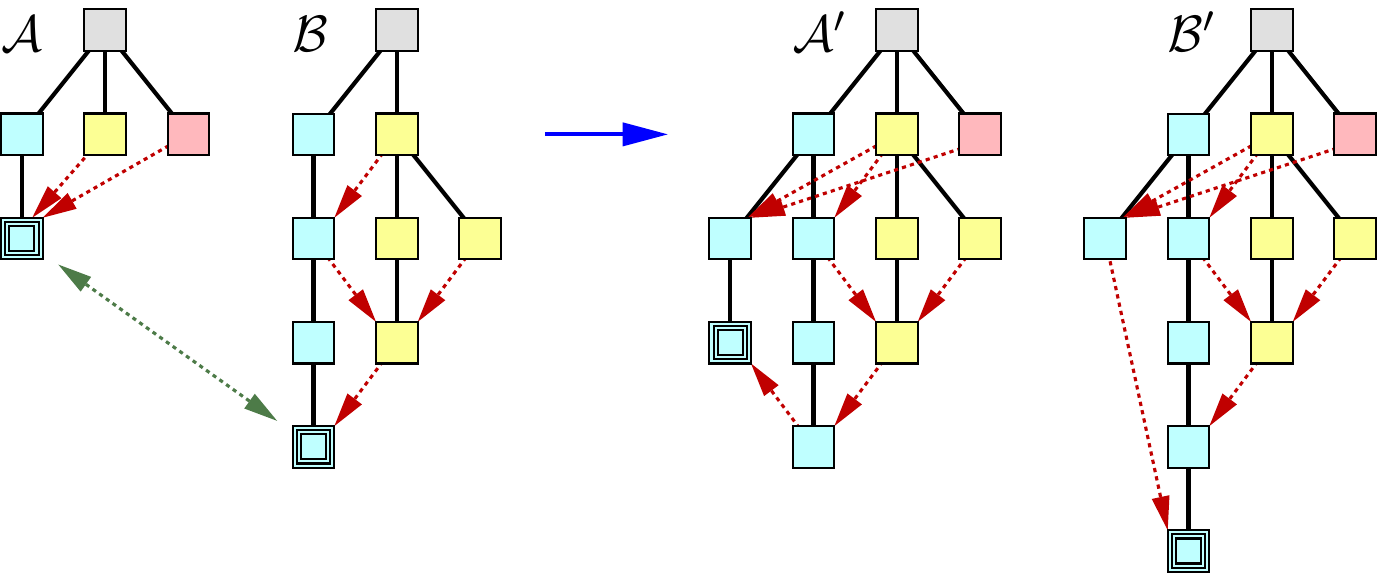}
\caption{Two agents in a semi-synchronous network share their views $\mathcal A$ and $\mathcal B$, which are updated as $\mathcal A'$ and $\mathcal B'$. Note that red edges may go upward or skip multiple levels. The highlighted nodes are the ``bottom nodes'' of their respective views, which can be identified as the only sinks (assuming black edges are directed away from the root).} \label{fig:ssynch}
\end{figure}

The concept of a round and the parameter $\tau$ are defined as in \cref{s:3.disc}. A notable distinction from the synchronous models discussed so far is that agents in a semi-synchronous network cannot reliably count communication steps, because they do not know for how many steps they have been inactive. Hence, two views shared by two agents may not have the same height, as the ones in \cref{fig:ssynch}.

\begin{figure}[t]
\centering
\includegraphics[scale=0.5]{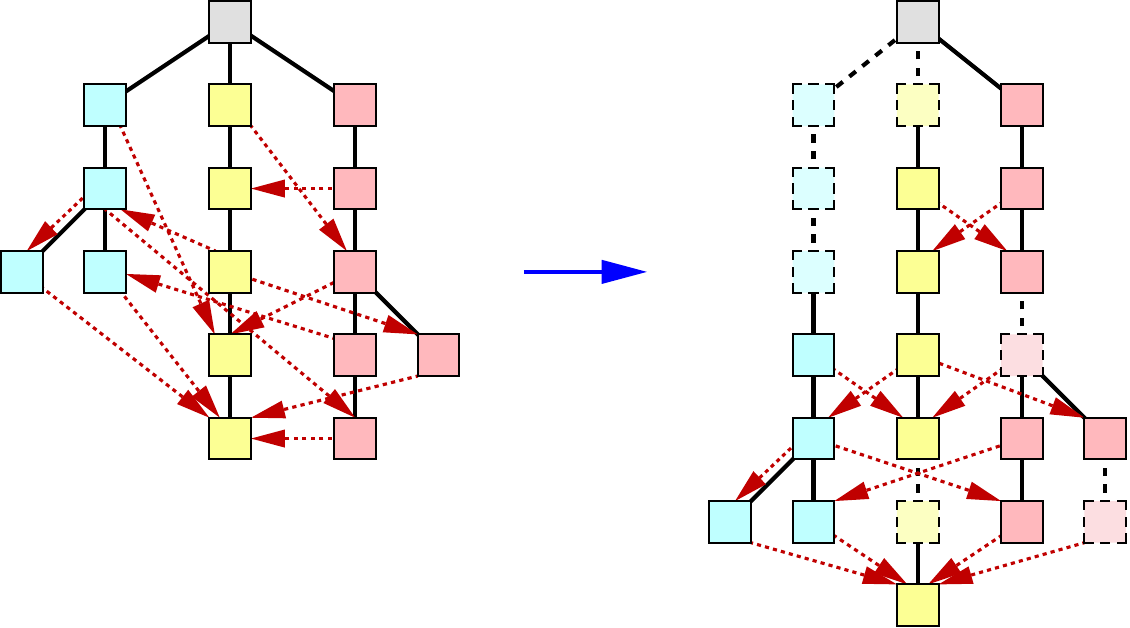}
\caption{Equalizing the view of an agent in a semi-synchronous network by adding dummy nodes. In the resulting view, every red edge connects a level to the next.} \label{fig:equal}
\end{figure}

Consequently, a red edge no longer has to connect a level to the next, but can span any number of levels. However, it is always possible to restore this property by adding \emph{dummy nodes} to represent inactive agents, resulting in an \emph{equalized} view (\cref{fig:equal}). Thus, we can reduce any semi-synchronous network to an equivalent synchronous one with the same parameter $\tau$ and apply the methods outlined in \cref{s:3.disc} to obtain algorithms with the same running times.

\subsection{Asynchronous Networks}\label{s:3.asynch}
A network is \emph{asynchronous} if messages can take an arbitrary, independent, and unpredictable amount of time to reach their destinations. Such a network is necessarily directed, and therefore some form of outdegree awareness is needed, as argued in \cref{s:3.directed}. A \emph{round} is now any minimal interval of time such that the messages that are sent and received form a strongly connected multigraph.

Observe that, even in asynchronous networks, time can always be discretized, assuming that no agent can send an infinite number of messages in a finite time.

Also, doing non-trivial computations with termination in asynchronous networks with no knowledge about the duration of a round is impossible.

However, there is a simple stabilization algorithm, as follows. When an agent sends some messages, it expands its view by adding a child to the bottom node and attaches its outdegree to the corresponding black edge, as in \cref{s:3.directed}. When it receives a multiset of messages, it updates its view as in \cref{s:3.ssynch}.

Then, the agent seeks the first interval of levels constituting a round where all nodes are non-branching. The outdegrees and red edges in this interval yield a system of equations that is solved as in the stabilizing algorithm of \cref{s:3.directed}.

Such a non-branching interval occurs in the history tree after at most $n-1$ rounds and appears in the agent's view in at most $n-1$ additional rounds. The total stabilization time is therefore $2n-2$ rounds, which is worst-case optimal.

\subsection{Port Awareness}\label{s:3.ports}
A widely studied scenario, especially within static networks, is when each agent has a local numbering for its incoming links (\emph{input port awareness}) or outgoing links (\emph{output port awareness}). It is evident that, although input port awareness enables each agent in a static network to identify all messages sent by the same neighbor, this feature has no clear meaning or effect in dynamic networks.

In contrast, output port awareness has a significant impact on both static and dynamic networks, as it not only implies outdegree awareness, but allows agents to assign a different tag to each message they send within a communication step. This is helpful in breaking network symmetry, thus facilitating certain computations. The history tree architecture for outdegree awareness of \cref{s:3.directed} can also be adapted to this model by simply attaching a port number to each red edge.

As an example, output port awareness combined with a unique leader makes the Counting problem particularly simple even in strongly connected directed dynamic networks. Indeed, all messages sent by the leader in a communication step have different tags, and so all agents that receive them become distinguishable. Generalizing, every node on a red path starting at a leader node must have an anonymity of~$1$. When a whole level $L_i$ in a view consists of such nodes, we can check if the outdegree of each node in $L_i$ matches the number of its outgoing red edges. If so, all agents have been accounted for, and we can return $n = |L_i|$.

This Counting algorithm terminates in $2n-1$ steps, greatly improving upon the one in \cref{s:3.term}. Observe that the lower bound in \cref{fig:lower} does not hold for networks with output port awareness, which may allow for even faster solutions.

\begin{tcolorbox}
{\bf Open Problem 6.} Can a Counting algorithm terminate in fewer than $2n-2$ steps in all strongly connected directed dynamic networks with output port awareness and a unique leader?
\end{tcolorbox}

A less obvious fact is that, in any non-branching level of the history tree of a strongly connected directed dynamic network with output port awareness, all nodes must have the same anonymity.\footnote{This observation, limited to static networks, is also found in~\cite[Theorems~3 and 10]{BV96}.} Indeed, if $A$ and $B$ are two classes of indistinguishable agents represented by nodes in a non-branching level, and an agent in $A$ receives a message tagged $k$ from an agent in $B$, then so do all agents in $A$ (or else their node would branch). Since no agent in $B$ can send more than one message tagged $k$, we have $|A|\leq |B|$. Repeating this reasoning for all pairs of communicating classes, and recalling that the network is strongly connected, we conclude that all classes must have the same size.

A remarkable consequence is that, with output port awareness, a unique leader can be elected if and only if it is possible to assign unique identifiers to all agents (also known as the \emph{Naming problem}). A terminating algorithm for both problems, if $n$ is known, is easily obtained by combining the Leader Election algorithm in \cref{s:3.leader} (which was designed for undirected networks only) with the stabilizing technique for directed networks in \cref{s:3.directed}.

\subsection{Varying Inputs}\label{s:3.inputs}
Suppose that agents have inputs that may change at every step. Note that the history tree architecture already supports attaching inputs to nodes, and so the basic stabilizing algorithms previously discussed can become \emph{streaming algorithms} that adaptively return the correct output with an amortized delay of $n-1$ steps.

\subsection{Self-Stabilization}\label{s:3.self}
A network algorithm is \emph{self-stabilizing} if it returns the correct output regardless of the initial state of the agents. This implies tolerance to memory corruption, transient faults, and agents dynamically joining and leaving the network.

Let us assume the network to be synchronous, directed, and dynamic. We remark that most existing self-stabilizing protocols, such as Boldi--Vigna's~\cite{BV02b}, are specifically designed for static networks, and cause agents to perpetually and incorrectly reset their states when executed in highly dynamic networks. 

We will give a \emph{universal} self-stabilizing protocol that constructs coherent views, allowing to convert any stabilizing algorithm into a self-stabilizing one (note that non-trivial terminating computations are impossible in this scenario).

\begin{figure}[t]
\centering
\includegraphics[scale=0.5]{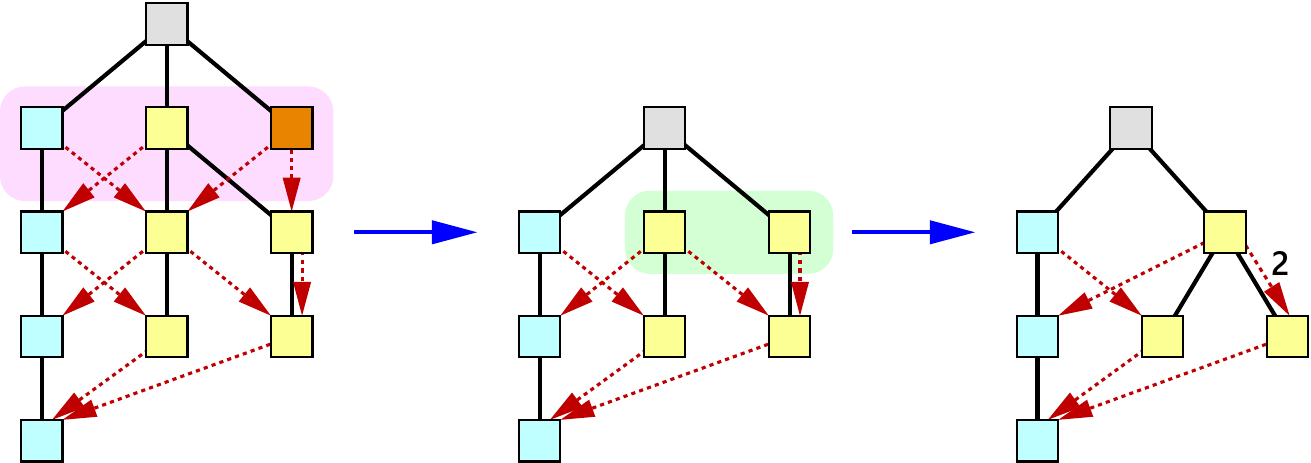}
\caption{A routine in the universal self-stabilizing protocol: delete level $L_0$ in the view, and then repeatedly merge nodes whose corresponding sub-views are isomorphic.} \label{fig:self}
\end{figure}

The core idea is that an agent can deliberately ``forget'' old information by deleting level $L_0$ of its current view, connect level $L_1$ to the root, and merge equivalent nodes from top to bottom to restore a well-formed view, as in \cref{fig:self}.

If the number of agents $n$ is known, they can simply update their views as usual (resetting their state if it does not encode a well-formed view) and delete old levels when their number exceeds some fixed threshold, e.g., $2n-2$. Eventually, this protocol produces views that correctly describe the last $2n-2$ communication steps, which are enough for the stabilizing algorithms of \cref{s:2.appl,s:3.directed}.

On the other hand, if $n$ is unknown, every agent updates its view as usual, but deletes level $L_0$ every \emph{two} steps. This is controlled by a local binary flag that is toggled at every step. However, before merging its view with a neighbor's view of different height, it deletes the top levels of the taller view to match the other. Also, if the taller view was its own, the agent does not toggle its flag for a step.

In $O(n)$ steps, all agents have views of equal height, as well as equal flags. Moreover, their views eventually describe an increasingly large number of previous communication steps, enabling any stabilizing algorithm to work correctly.

\begin{tcolorbox}
{\bf Open Problem 7.} Is there a universal self-stabilizing protocol for \emph{semi-synchronous} strongly connected directed dynamic networks?
\end{tcolorbox}

\subsection{Finite-State Stabilization}\label{s:3.finite}
The stabilizing algorithms described so far assume that all agents constantly update their views at every step, which requires an unlimited amount of internal memory. To mitigate this, we will now give a \emph{universal finite-state} protocol that enables the conversion of any stabilizing algorithm into one that uses a finite amount of memory (as a function of $n$), albeit with an extra delay.

Let us assume the network to be synchronous, undirected, connected, and dynamic. A level $L_i$ in a view is said to be \emph{suitable} if every node in $L_{i-1}$ has a unique child in $L_i$ and the red edges between these two levels are compatible with a connected network. Recall that the basic stabilizing algorithms of \cref{s:2.appl} do all computations using only the shallowest suitable level of a view.

Now, when two neighboring agents share views whose shallowest suitable levels are isomorphic, they do not merge their views. Essentially, this removes a link from the communication network for that step. Also, when an agent receives no views that it can merge, it skips updating its view altogether, i.e., it remains inactive for that step. Note that this makes the network semi-synchronous, so views must be \emph{equalized} prior to being used by the protocol (see \cref{s:3.ssynch}).

According to this protocol, if all agents remain inactive forever, they all share the same suitable level, which is enough to perform correct computations. Otherwise, some inactive agent acquires relevant information within $n-1$ steps and reactivates. Moreover, since there are at most $n-1$ branching nodes in the history tree, the agents can incorrectly guess a suitable level at most $n-1$ times. Thus, the protocol is finite-state, but introduces an overhead of $O(n^2)$ steps.

\begin{tcolorbox}
{\bf Open Problem 8.} Is there a universal finite-state stabilizing protocol for connected undirected dynamic networks with an overhead of $O(n)$ steps?
\end{tcolorbox}

Another open problem is to design a protocol that is both self-stabilizing and finite-state. In \cref{s:3.self} we already gave one, but it assumes $n$ to be known.

\begin{tcolorbox}
{\bf Open Problem 9.} Is there a universal finite-state self-stabilizing protocol for connected undirected dynamic networks of unknown size?
\end{tcolorbox}

\subsection{Memoryless Computation}\label{s:3.memoryless}
An agent is \emph{memoryless} if its state is reset at every communication step, meaning its entire memory is erased after it sends messages and before receiving messages from neighbors. The goal is to design a universal protocol that enables memoryless agents to construct coherent views of \emph{some} history tree related to the network. This would allow basic algorithms to run correctly within the memoryless model.

\begin{tcolorbox}
{\bf Open Problem 10.} Under what assumptions is there a universal memoryless protocol for anonymous networks?
\end{tcolorbox}

\subsection{Congested Networks}\label{s:3.cong}
In a \emph{congested network} of $n$ agents, communication links have a logarithmic bandwidth, and therefore the size of each message is limited to $O(\log n)$ bits.\footnote{Agents do not necessarily know this limit, since they generally do not know $n$. Nonetheless, if they attempt to send larger messages, the network's behavior is undefined. Ideally, algorithms should be designed to automatically avoid this situation.}

It is evident that the basic technique of encoding a view as a single message is not applicable in a congested network, because the size of a view grows polynomially at every step (see \cref{s:2.appl}). Furthermore, any naive approach that simply subdivides a view into smaller pieces to be transmitted over multiple steps is clearly ineffective in anonymous dynamic networks.

Correct protocols for constructing views in congested dynamic networks are found in~\cite{DV23}. If the dynamic diameter $d$ is known, then $O(\log n)$ bits of information can be reliably \emph{broadcast} in phases of $d$ steps. This makes it possible for agents to share enough information to construct a history tree that is compatible with their network, one level at a time. Every node in this history tree has a unique label of logarithmic size, which also identifies all agents represented by that node. This protocol can be used to solve the Counting problem in $O(dn^2)$ steps.

If $d$ is unknown but there is a unique leader, there is a more complex protocol that attempts to estimate $d$ by implementing a \emph{reset module} that repeatedly doubles the estimate every time a broadcasting error is detected. The leader coordinates the process, ensuring that all agents agree on the same version of the history tree. This yields a Counting algorithm that terminates in $O(n^3)$ steps.

\begin{tcolorbox}
{\bf Open Problem 11.} Is there a terminating Counting algorithm for congested dynamic networks \emph{with more than one leader}?
\end{tcolorbox}

It was proved in~\cite{DPR13} that solving the Counting problem in congested networks by broadcasting ``tokens'' requires at least $\Omega(n^2/\log n)$ steps. On the other hand, there is a solution in $O(n^2)$ steps if messages have size $O(n\log n)$ bits~\cite{DV23}.

\begin{tcolorbox}
{\bf Open Problem 12.} Can a Counting algorithm terminate in $O(n^2)$ steps in all congested dynamic networks with a unique leader?
\end{tcolorbox}

We find it fitting to conclude this note with an open problem that is unlikely to have a solution using history trees.

\begin{tcolorbox}
{\bf Open Problem 13.} Is there a terminating Counting algorithm for congested dynamic networks where agents have \emph{logarithmic-sized memory}?
\end{tcolorbox}

\begin{credits}
\subsubsection{\ackname} The author expresses profound gratitude to collaborators J\'er\'emie Chalopin, Giuseppe A.\ Di Luna, and Federico Poloni for numerous inspiring discussions. This work was supported by the JSPS KAKENHI grant 23K10985, the University of Aizu Competitive Research Funding for FY~2023, and the University of Salerno.

\subsubsection{\discintname}
The author has no competing interests to declare that are
relevant to the content of this article.
\end{credits}
%
%
%
\bibliographystyle{splncs04}
\bibliography{bibliography}
\end{document}